\documentclass[12pt,a4paper]{article}

\usepackage{epsfig}
\usepackage{amsmath,amsfonts,amssymb}
\usepackage{t1enc}
\usepackage{cite}
\usepackage{colortbl}
\usepackage{pifont}

\providecommand{\openone}{\leavevmode\hbox{\small1\kern-3.8pt\normalsize1}}

\newcommand{\RE}{\text{Re}\,}

\newcommand{\bmu}{\mathcal{B}_\mu}
\newcommand{\wmu}{\mathcal{W}_\mu}
\newcommand{\gmu}{\mathcal{G}_\mu}
\newcommand{\hmu}{\mathcal{H}_\mu}
\newcommand{\qmu}{\mathcal{Q}_\mu^5}
\newcommand{\ymu}{\mathcal{Y}_\mu^5}
\newcommand{\qmuo}{\mathcal{Q}_\mu^1}
\newcommand{\ymuo}{\mathcal{Y}_\mu^1}
\newcommand{\omf}{\omega^4}
\newcommand{\OMf}{\Omega^4}
\newcommand{\omo}{\omega^1}
\newcommand{\OMo}{\Omega^1}
\newcommand{\so}{\sigma}
\newcommand{\So}{\Sigma}

\newcommand{\afb}{A_\text{FB}}

\newcommand{\oh}{\textstyle \frac{1}{2}}
\newcommand{\ot}{\textstyle \frac{1}{3}}
\newcommand{\of}{\textstyle \frac{1}{4}}
\newcommand{\os}{\textstyle \frac{1}{6}}
\newcommand{\oei}{\textstyle \frac{1}{8}}
\newcommand{\otw}{\textstyle \frac{1}{12}}

\newcommand{\gM}{\gamma^\mu}
\newcommand{\gm}{\gamma_\mu}
\newcommand{\la}{\lambda^a}
\newcommand{\tI}{\tau^I}
\newcommand{\pr}{\Pi}

\newcommand{\Cqq}{C_{qq}}
\newcommand{\Cqqp}{C_{qq'}}
\newcommand{\Cuu}{C_{uu}}
\newcommand{\Cqu}{C_{qu}}
\newcommand{\Cqup}{C_{qu'}}
\newcommand{\Cqd}{C_{qd}}
\newcommand{\Cqdp}{C_{qd'}}
\newcommand{\Cud}{C_{ud}}
\newcommand{\Cudp}{C_{ud'}}
\newcommand{\Cqqe}{C_{qq\epsilon}}
\newcommand{\Cqqep}{C_{qq\epsilon'}}

\begin{document}

\begin{center}
\begin{Large}
{\bf Probing the Tevatron $t \bar t$ asymmetry at LHC}
\end{Large}

\vspace{0.5cm}
J. A. Aguilar--Saavedra, M. P\'erez-Victoria \\[0.2cm] 
{\it Departamento de F\'{\i}sica Te\'orica y del Cosmos and CAFPE, \\
Universidad de Granada, E-18071 Granada, Spain}
\end{center}

\begin{abstract}
We use an effective operator framework to study the contributions to the Tevatron $t \bar t$ asymmetry from arbitrary vector bosons and scalars, and compare with their effect on the $t \bar t$ tail at LHC. Our study shows, for example, that models reproducing the $t \bar t$ asymmetry by exchange of $Z'$ and $W'$ bosons or colour-triplet scalars lead to a large enhancement in the $t \bar t$ tail at LHC. This fact can be used to exclude these models as the sole explanation for the asymmetry, using the data already collected by CMS and ATLAS.
Our analysis is model independent in the sense that we scan over all possible extra particles contributing to the asymmetry, and allow for general couplings.
We also explore a class of Standard Model extensions which can accommodate the Tevatron asymmetry without contributing to the total $t \bar t$ cross section at first order, so that the enhancement of the tail at Tevatron and LHC is moderate.

\end{abstract}

\section{Introduction} 

The measurement of the $t \bar t$ forward-backward (FB) asymmetry at Tevatron~\cite{:2007qb,Aaltonen:2008hc,Aaltonen:2011kc} has motivated a plethora of models which attempt to accommodate the experimental values, up to $3.4\sigma$ larger than the prediction of the Standard Model (SM). This task is not straightforward because the measured $t \bar t$ cross section is in good agreement with the SM: any ``generic'' addition to the $t \bar t$ production amplitude, large enough to produce the observed FB asymmetry, will easily give rise to too large a departure in the total rate. Many of the proposed models circumvent this problem at the expense of a cancellation between (linear) interference and (quadratic) new physics terms in the total cross section,
\begin{equation}
\sigma(t\bar t) = \sigma_\text{SM} + \delta \sigma_\text{int} + \delta \sigma_\text{quad} \,,
\end{equation}
where $\sigma_\text{SM}$ is the SM cross section, $\delta \sigma_\text{quad}$ the one corresponding to the new physics and $\delta \sigma_\text{int}$ the interference term. 
The cancellation $\delta \sigma_\text{int} + \delta \sigma_\text{quad} \simeq 0$ requires a new large amplitude $A_\text{new} \sim - 2 A_\text{SM}$ which, obviously, should have observable effects elsewhere. The ideal candidate to search for these effects is the Large Hadron Collider (LHC).

New physics in $u \bar u,d \bar d \to t \bar t$ which produces such a large cancellation in the Tevatron cross section will likely produce an observable enhancement in the $t \bar t$ tail at LHC, even if at this collider top pair production is dominated by gluon fusion. (The tail is also enhanced at Tevatron energies but in the majority of the proposed models  the deviations are compatible with present measurements~\cite{Aaltonen:2011kc}.) In some cases, this effect should be visible  already with the data collected in 2010. Conversely, if large deviations are not reported, a number of candidates to explain the Tevatron $t \bar t$ asymmetry will be excluded from the list.

In this paper we make these arguments quantitative for a wide class of SM extensions. We consider general new vector bosons and scalars, classified by their transformation properties under the SM gauge group $\text{SU}(3) \times \text{SU}(2)_L \times \text{U}(1)_Y$, and study their possible effects in $t \bar t$ production. We use effective field theory to consistently (i) integrate out the new heavy states and obtain their contribution to $u \bar u,d \bar d \to t \bar t$ in terms of four-fermion operators; (ii) obtain the cross sections in terms of effective operator coefficients. This allows us to find, for each vector boson and scalar representation, a relation between the possible values of the asymmetry $\afb$ at Tevatron and the excess in the $t \bar t$ tail at LHC. The discussion of all possible vector boson and scalar representations within the model-independent effective operator framework allows us to make stronger statements than in other model-independent studies~\cite{Zhang:2010dr,Degrande:2010kt,Blum:2011up,Delaunay:2011gv}. At the same time, any model with several vector bosons and scalars can be considered in our framework by simply summing the effective operator coefficients corresponding to the integration of each new particle. Previous studies of LHC signals associated to the FB asymmetry within particular models have been presented in
Refs.~\cite{Dorsner:2009mq,Jung:2010yn,Choudhury:2010cd,Cao:2011ew,
Bai:2011ed,Berger:2011ua,Bhattacherjee:2011nr,Patel:2011eh,Gresham:2011dg}.

After this analysis, we explore an alternative way to produce a large asymmetry with moderate effects in the $t \bar t$ tail at LHC. The key for this mechanism is the observation that one can also obtain a large $\afb$ without significant changes in the total $t \bar t$ cross section by introducing new physics which only contributes to the latter at quadratic order. This has been studied before, in the effective formalism, in Ref.~\cite{Degrande:2010kt}. If we write
\begin{eqnarray}
\sigma^F (t \bar t) & = & \sigma_\text{SM} ^F +  \sigma_\text{int}^F + \sigma_\text{quad}^F \,, \notag \\
\sigma^B  (t \bar t) & = & \sigma_\text{SM} ^B +  \sigma_\text{int}^B + \sigma_\text{quad}^B \,,
\end{eqnarray}
for the forward (F) and backward (B) cross sections, the total rate is maintained at first order provided $\sigma_\text{int}^F + \sigma_\text{int}^B \simeq 0$, which can be achieved, for example, with a new vector boson and a scalar. For models fulfilling this cancellation the size of the new physics contributions required to accommodate the experimental value of $\afb$ are smaller. Therefore, these models provide a better agreement of the $t \bar t$ tail with Tevatron measurements, and predict a much smaller tail at LHC, which is still potentially observable with forthcoming measurements.

We remark that the use of effective field theory (with the assumption that the new physics is too heavy to be directly produced at LHC) does not limit much the generality of our conclusions. For $t$-channel exchange of new vector bosons or scalars, integrating out the new particles gives a good estimate for masses $M \gtrsim 1$ TeV. For $s$-channel exchange this approximation is worse, but the cross section enhancement produced by the new particle(s) is always larger than the one from the corresponding four-fermion operator(s), and in this sense our predictions are conservative. On the other hand, if new physics in the $t \bar t$ tail is not seen at LHC, the new resonances are heavy and the effective operator framework
can be safely used. It is also worth pointing out that we include $1/\Lambda^4$ corrections arising from the quadratic terms in new physics in the cross sections. The contributions from the interference of $1/\Lambda^4$ operators with the SM can be neglected, as they are suppressed with respect to the former for the values of parameters that are required to explain the $t \bar t$ asymmetry.

\section{Extra bosons, operators and $t \bar t$ production}
\label{sec:2}

There are ten possible $\text{SU}(3) \times \text{SU}(2)_L \times \text{U}(1)_Y$ representations~\cite{delAguila:2010mx} for new vector bosons contributing to $u \bar u,d \bar d \to t \bar t$, while for scalars eight representations contribute. They are collected in Table~\ref{tab:lagr}, where the first column indicates the label used to refer to them.\footnote{We note that for $\wmu$ and $\hmu$ the normalisation in the Lagrangian differs from Ref.~\cite{delAguila:2010mx} by a factor of two, to simplify the presentation of the limits.}
The relevant interaction Lagrangian is included as well, indicating the symmetry properties, if any, of the coupling matrices $g_{ij}$. We use standard notation with left-handed doublets $q_{Li}$ and right-handed singlets $u_{Ri}$, $d_{Ri}$; $\tI$ are the Pauli matrices, $\la$ the Gell-Mann matrices normalised to $\text{tr}(\la \lambda^b) = 2 \delta_{ab}$ and $\tilde \phi = \epsilon \phi$, $\psi^c = C \bar \psi^T$, with $\epsilon=i\tau^2$ and $C$ the charge conjugation matrix. The indices $a,b,c$ denote colour, and $\varepsilon_{abc}$ is the totally antisymmetric tensor.

\begin{table}[p]
\begin{center}
\begin{tabular}{c|clc}
Label & Rep. & \multicolumn{1}{c}{Interaction Lagrangian} & Sym. \\
\hline
$\bmu$ & $(1,1)_0$ 
  & $-\left( g_{ij}^q \bar q_{Li} \gM q_{Lj} 
  + g_{ij}^u \bar u_{Ri} \gM u_{Rj} 
  + g_{ij}^d \bar d_{Ri} \gM d_{Rj} \right) \bmu $ & $g=g^\dagger$ \\[1mm]
$\wmu$ & $(1,3)_0$
  & $- g_{ij} \bar q_{Li}  \gM \tau^I q_{Lj} \, \mathcal{W}_\mu^I$
  & $g=g^\dagger$ \\[1mm]
$\bmu^1$ & $(1,1)_1$ 
  & $- g_{ij} \bar d_{Ri} \gM u_{Rj} \, \bmu^{1\dagger} + \text{h.c.}$ & -- \\[1mm]
$\gmu$ & $(8,1)_0$
  & $- \left( g_{ij}^q \bar q_{Li} \gM \frac{\la}{2} q_{Lj} 
  + g_{ij}^u \bar u_{Ri} \gM \frac{\la}{2} u_{Rj} 
  + g_{ij}^d \bar d_{Ri} \gM \frac{\la}{2} d_{Rj} \right) \mathcal{G}_\mu^a$ & $g=g^\dagger$ \\[1mm]
$\hmu$ & $(8,3)_0$
  & $- g_{ij} \bar q_{Li}  \gM \tau^I \frac{\la}{2} q_{Lj} \, \mathcal{H}_\mu^{aI}$ & $g=g^\dagger$ \\[1mm]
$\gmu^1$ & $(8,1)_1$ 
  & $- g_{ij} \bar d_{Ri} \gM \frac{\la}{2} u_{Rj} \, \gmu^{1a\dagger} + \text{h.c.}$ & -- \\[1mm]
$\qmuo$ & $(3,2)_{\frac{1}{6}}$
  & $-g_{ij} \varepsilon_{abc} \bar d_{Rib} \gM \epsilon q_{Ljc}^c \, \mathcal{Q}_\mu^{1a\dagger} + \text{h.c.}$ & -- \\[1mm]
$\qmu$ & $(3,2)_{-\frac{5}{6}}$
  & $-g_{ij} \varepsilon_{abc} \bar u_{Rib} \gM \epsilon q_{Ljc}^c \, \mathcal{Q}_\mu^{5a\dagger} + \text{h.c.}$ & -- \\[1mm]
$\ymuo$ & $(\bar 6,2)_{\frac{1}{6}}$
  & $-g_{ij} \oh \left[ \bar d_{Ria} \gM \epsilon q_{Ljb}^c + 
  \bar d_{Rib} \gM \epsilon q_{Lja}^c \right] \mathcal{Y}_\mu^{1ab\dagger}  + \text{h.c.}$ & -- \\[1mm]
$\ymu$ & $(\bar 6,2)_{-\frac{5}{6}}$
  & $-g_{ij} \oh \left[ \bar u_{Ria} \gM \epsilon q_{Ljb}^c + 
  \bar u_{Rib} \gM \epsilon q_{Lja}^c \right] \mathcal{Y}_\mu^{5ab\dagger}  + \text{h.c.}$ & -- \\[1mm]
$\phi$ & $(1,2)_{-\frac{1}{2}}$
  & $- g_{ij}^u \bar q_{Li} u_{Rj} \, \phi - g_{ij}^d \bar q_{Li} d_{Rj} \, \tilde \phi  + \text{h.c.}$ & -- \\[1mm]
%f
$\Phi$ & $(8,2)_{-\frac{1}{2}}$
  & $- g_{ij}^u \bar q_{Li} \frac{\la}{2} u_{Rj} \, \Phi^a - g_{ij}^d \bar q_{Li} \frac{\la}{2} d_{Rj} \, \tilde \Phi^a  + \text{h.c.}$ & -- \\[1mm]
$\omo$ & $(3,1)_{-\frac{1}{3}}$
  & $- g_{ij} \varepsilon_{abc} \bar d_{Rib} u_{Rjc}^c \, \omega^{1a\dagger} + \text{h.c.}$ & -- \\[1mm]
$\OMo$ & $(\bar 6,1)_{-\frac{1}{3}}$
  & $-g_{ij} \oh \left[ \bar d_{Ria} u_{Rjb}^c + 
  \bar d_{Rib} u_{Rja}^c \right] \Omega^{1ab\dagger} + \text{h.c.}$ & -- \\[1mm]
$\omf$ & $(3,1)_{-\frac{4}{3}}$
  & $- g_{ij} \varepsilon_{abc} \bar u_{Rib} u_{Rjc}^c \, \omega^{4a\dagger} + \text{h.c.}$ & $g=-g^T$ \\[1mm]
$\OMf$ & $(\bar 6,1)_{-\frac{4}{3}}$
  & $-g_{ij} \oh \left[ \bar u_{Ria} u_{Rjb}^c + 
  \bar u_{Rib} u_{Rja}^c \right] \Omega^{4ab\dagger} + \text{h.c.}$ & $g=g^T$ \\[1mm]
$\so$ & $(3,3)_{-\frac{1}{3}}$
  & $- g_{ij} \varepsilon_{abc} \bar q_{Lib} \tI \epsilon q_{Ljc}^c \, \sigma^{a\dagger} + \text{h.c.}$ & $g=-g^T$ \\[1mm]
$\So$ & $(\bar 6,3)_{-\frac{1}{3}}$
  & $-g_{ij} \oh \left[ \bar q_{Lia} \tI \epsilon q_{Ljb}^c + 
  \bar q_{Lib} \tI \epsilon q_{Lja}^c \right] \So^{Iab\dagger} + \text{h.c.}$ & $g=g^T$
\end{tabular}
\end{center}
\caption{Vector bosons and scalar representations mediating $u \bar u,d \bar d \to t \bar t$.}
\label{tab:lagr}
\end{table}

\begin{table}[p]
\begin{center}
\begin{small}
\begin{tabular}{c|ccccccc}
& $C_{qq}^{3113}$  & $C_{qq'}^{1133}$ & $C_{uu}^{3113}$
& $C_{ud'}^{3311}$ & $C_{qu}^{1331}$  & $C_{qu}^{3113}$ 
& $C_{qd}^{3113}$ \\[1mm]
\hline \\[-4mm]
$\bmu$ & $-|g_{13}^q|^2$ & -- & $-|g_{13}^u|^2$ & -- & -- & -- & -- \\[1mm]
$\wmu$ & $|g_{13}|^2$ & $-2 |g_{13}|^2$ & -- & -- & -- & -- & -- \\[1mm]
$\gmu$ & $\os |g_{13}^q|^2$ & $-\oh g_{11}^q g_{33}^q$ 
& \!\!\begin{tabular}{c}$\os |g_{13}^u|^2$ \\ $-\oh g_{11}^u g_{33}^u$
\end{tabular}\!\!
& $-\of g_{33}^u g_{11}^d$ & $\oh g_{11}^q g_{33}^u$ & $\oh g_{33}^q g_{11}^u$ & $\oh g_{33}^q g_{11}^d$ \\[1mm]
$\hmu$ 
& \!\!\begin{tabular}{c}$-\os |g_{13}|^2$ \\ $- g_{11} g_{33}$
\end{tabular}\!\!
& \!\!\begin{tabular}{c}$\ot |g_{13}|^2$ \\ $+\oh g_{11} g_{33}$
\end{tabular}\!\!
& -- & -- & -- & -- & -- \\[1mm]
$\bmu^1$ & -- & -- & -- & $-\oh|g_{13}|^2$ & -- & -- & -- \\[1mm]
$\gmu^1$ & -- & -- & -- & $\otw |g_{13}|^2$ & -- & -- & -- \\[1mm]
$\qmuo$ & -- & -- & -- & -- & -- & -- & $|g_{13}|^2$ \\[1mm]
$\qmu$ & -- & -- & -- & -- & $|g_{31}|^2$ & $|g_{13}|^2$ & -- \\[1mm]
$\ymuo$ & -- & -- & -- & -- & -- & -- & $-\oh |g_{13}|^2$ \\[1mm]
$\ymu$ & -- & -- & -- & -- & $-\oh|g_{31}|^2$ & $-\oh|g_{13}|^2$\\[1mm]
$\phi$ & -- & -- & -- & -- & $\oh |g_{13}^u|^2$ & $\oh |g_{31}^u|^2$ & $\oh |g_{31}^d|^2$\\[1mm]
$\Phi$ & -- & -- & -- & -- & $-\otw |g_{13}^u|^2$ & $-\otw |g_{31}^u|^2$ & $-\otw |g_{31}^d|^2$\\[1mm]
$\omo$ & -- & -- & -- & $-\of |g_{13}|^2$ & -- & -- & -- \\[1mm]
$\OMo$ & -- & -- & -- & $\oei |g_{13}|^2$ & -- & -- & -- \\[1mm]
$\omf$ & -- & -- & $-2|g_{13}|^2$ & -- & -- & -- & -- \\[1mm]
$\OMf$ & -- & -- & $|g_{13}|^2$ & -- & -- & -- & -- \\[1mm]
$\so$ & $-2|g_{13}|^2$ & $-2|g_{13}|^2$ & -- & -- & -- & -- & -- \\[1mm]
$\So$ & $|g_{13}|^2$ & $|g_{13}|^2$ & -- & -- & -- & -- & --
\end{tabular}
\end{small}
\end{center}
\caption{Coefficients of effective operators interfering with the SM amplitudes for $u \bar u,d\bar d \to t \bar t$. The new physics scale $\Lambda$ equals the mass of the new particle or multiplet.}
\label{tab:CAint}
\end{table}

\begin{table}[p]
\begin{center}
\begin{tabular}{c|ccccccc}
& $C_{qq}^{1133}$ & $C_{qq'}^{3113}$ & $C_{uu}^{1133}$
& $C_{ud}^{3311}$ \\[1mm]
\hline \\[-4mm]
$\bmu$ & $-g_{11}^q g_{33}^q$ & -- & $-g_{11}^u g_{33}^u$ & $-\oh g_{33}^u g_{11}^d$ \\[1mm]
$\wmu$ & $g_{11}^q g_{33}^q$ & $-2 g_{11}^q g_{33}^q$ & -- & -- \\[1mm]
$\gmu$ & $\os g_{11}^q g_{33}^q$ & $-\oh |g_{13}^q|^2$ 
& \!\!\begin{tabular}{c}$\os g_{11}^u g_{33}^u$ \\ $-\oh |g_{13}^u|^2$
\end{tabular}
& $\otw g_{33}^u g_{11}^d$ \\[1mm]
$\hmu$
& \!\!\begin{tabular}{c}$-\os g_{11} g_{33}$ \\ $- |g_{13}|^2$
\end{tabular}
& \!\!\begin{tabular}{c}$\oh |g_{13}|^2$ \\ $+\ot g_{11} g_{33}$
\end{tabular}
& -- & -- \\[1mm]
$\gmu^1$ & -- & -- & -- & $-\of |g_{13}|^2$ \\[1mm]
%
%$\qmu$ & -- & -- & -- & -- \\[1mm]
%
%$\ymu$ & -- & -- & -- & -- \\[1mm]
%
%$\phi$ & -- & -- & -- & -- \\[1mm]
%
%$\Phi$ & -- & -- & -- & -- \\[1mm]
%
$\omo$ & -- & -- & -- & $\of |g_{13}|^2$ \\[1mm]
$\OMo$ & -- & -- & -- & $\oei |g_{13}|^2$ \\[1mm]
$\omf$ & -- & -- & $2|g_{13}|^2$ & -- \\[1mm]
$\OMf$ & -- & -- & $|g_{13}|^2$ & -- \\[1mm]
$\so$ & $2|g_{13}|^2$ & $2|g_{13}|^2$ & -- & -- \\[1mm]
$\So$ & $|g_{13}|^2$ & $|g_{13}|^2$ & -- & --
\end{tabular}
\end{center}
\caption{Coefficients of $\bar L L \bar L L$ and $\bar R R \bar R R$ effective operators contributing to $u \bar u,d\bar d \to t \bar t$ only at quadratic level. For the representations not listed all these coefficients vanish. The new physics scale $\Lambda$ equals the mass of the new particle or multiplet.}
\label{tab:CA1}
\end{table}

\begin{table}[p]
\begin{center}
\begin{tabular}{c|ccccccc}
& $C_{qu}^{3311}$ & $C_{qu'}^{1331}$ & $C_{qu'}^{3113}$ & $C_{qu'}^{3311}$
& $C_{qd'}^{3113}$ \\[1mm]
\hline \\[-4mm]
$\bmu$ & -- & $g_{11}^q g_{33}^u$ & $g_{33}^q g_{11}^u$ & $2 g_{13}^{q*} g_{13}^u$ & $g_{33}^q g_{11}^d$ \\[1mm]
%
%$\wmu$ & -- & -- & -- & -- & -- \\[1mm]
%
$\gmu$ & $g_{13}^{q*} g_{13}^u$ & $-\os g_{11}^q g_{33}^u$ & $-\os g_{33}^q g_{11}^u$ & $-\ot g_{13}^{q*} g_{13}^u$ & $-\os g_{33}^q g_{11}^d$ \\[1mm]
%
%$\hmu$ & -- & -- & -- & -- & -- \\[1mm]
%
$\qmuo$ & -- & -- & -- & -- & $-|g_{13}|^2$ \\[1mm]
$\qmu$ & $2 g_{13} g_{31}^*$ & $-|g_{31}|^2$ & $-|g_{13}|^2$ &
$-2 g_{13} g_{31}^*$ & -- \\[1mm]
$\ymuo$ & -- & -- & -- & -- & $-\oh |g_{13}|^2$ \\[1mm]
$\ymu$  & $- g_{13} g_{31}^*$ & $-\oh |g_{31}|^2$ & $-\oh |g_{13}|^2$ & 
$-g_{13} g_{31}^*$ & -- \\[1mm]
$\phi$ & $g_{11}^{u*} g_{33}^u$ & -- & -- & -- & -- \\[1mm]
$\Phi$ & $-\os g_{11}^{u*} g_{33}^u$ & $\of |g_{13}^u|^2$ & $\of |g_{31}^u|^2$ & $\oh g_{11}^{u*} g_{33}^u$ & $\of |g_{31}^d|^2$ \\[1mm]
%
%$\OMf$ & -- & -- & -- & -- & -- \\[1mm]
%
%$\So$ & -- & -- & -- & -- & -- 
\end{tabular}
\end{center}
\caption{Coefficients of $\bar L R \bar R L$ effective operators contributing to $u \bar u,d\bar d \to t \bar t$ only at quadratic level. For the representations not listed all these coefficients vanish. The new physics scale $\Lambda$ equals the mass of the new particle or multiplet.}
\label{tab:CA2}
\end{table}

\begin{table}[htb]
\begin{center}
\begin{tabular}{c|ccccccc}
& $C_{qq\epsilon}^{1331}$ & $C_{qq\epsilon}^{3311}$ 
& $C_{qq\epsilon'}^{1331}$ & $C_{qq\epsilon'}^{3311}$ \\
\hline \\[-4mm]
%$\bmu$ & -- & -- & -- & -- \\[1mm]
%
%$\wmu$ & -- & -- & -- & -- \\[1mm]
%
%$\gmu$ & -- & -- & -- & -- \\[1mm]
%
%$\hmu$ & -- & -- & -- & -- \\[1mm]
%
%$\qmu$ & -- & -- & -- & -- \\[1mm]
%
%$\ymu$ & -- & -- & -- & -- \\[1mm]
%
$\phi$ & $g_{13}^u g_{31}^d$ & $g_{33}^u g_{11}^d$ & -- & -- \\[1mm]
$\Phi$ & $-\os g_{13}^u g_{31}^d$ & $-\os g_{33}^u g_{11}^d$ & $\oh g_{13}^u g_{31}^d$ & $\oh g_{33}^u g_{11}^d$ \\[1mm]
%
%$\OMf$ & -- & -- & -- & -- \\[1mm]
%
%$\So$ & -- & -- & -- & --
\end{tabular}
\end{center}
\caption{Coefficients of $\bar L R \bar L R$ effective operators contributing to $u \bar u,d\bar d \to t \bar t$ only at quadratic level. For the representations not listed all these coefficients vanish. The new physics scale $\Lambda$ equals the mass of the new particle or multiplet.}
\label{tab:CA3}
\end{table}

The four-fermion operators contributing to $t\bar t$ production, including those that do not interfere with the SM QCD amplitude and only appear at quadratic level, have been given in Ref.~\cite{AguilarSaavedra:2010zi}. (The operators which interfere with the SM were given in Ref.~\cite{Jung:2009pi}.)
We collect in Tables~\ref{tab:CAint}--\ref{tab:CA3} the values of the corresponding coefficients for all the vector and scalar irreducible representations that can induce these operators. For vector bosons the coefficients have previously been obtained in Ref.~\cite{delAguila:2010mx}. The notation for four-fermion operators is given in appendix~\ref{sec:a}. We stress again that including
both interference $1/\Lambda^2$ and quadratic $1/\Lambda^4$ terms in our calculations is not inconsistent, despite the fact that we are not considering dimension-eight operators. When quadratic terms are relevant for $t \bar t$ production (large couplings) the missing dimension-eight terms are sub-leading in the classes of SM extensions we consider. A related discussion about the importance of $1/\Lambda^2$ and $1/\Lambda^4$ contributions from dimension-six operators has been presented in Ref.~\cite{AguilarSaavedra:2010sq}.

We evaluate the FB asymmetry
\begin{equation}
A_\text{FB} = \frac{\sigma^F-\sigma^B}{\sigma^F+\sigma^B} =
\frac{\sigma^F_\text{SM}+ \delta \sigma^F -\sigma_\text{SM}^B- \delta \sigma^B}{\sigma^F_\text{SM}+\delta \sigma^F +\sigma_\text{SM}^B +\delta \sigma^B} \,,
\end{equation}
using the SM predictions~\cite{Campbell:1999ah}
\begin{align}
& A_\text{FB}^\text{SM} =  0.058 \pm 0.009 && \text{(inclusive)} \,, \notag \\
& A_\text{FB}^\text{SM} =  0.088 \pm 0.013 && (m_{t\bar t} > 450~\text{GeV}) \,.
\label{ec:afbSM}
\end{align}
and the new contributions from four-fermion operators $\delta \sigma^{F,B}$, parameterised in terms of effective operator coefficients and numerical constants. The explicit expressions are collected in appendix~\ref{sec:a}. It is important to point out that positive operator coefficients always increase $\afb$ at first order ($1/\Lambda^2$ interference with the SM), as it follows from Eqs.~(\ref{ec:int}).
We choose, among the recently reported measurements of the FB asymmetry~\cite{Aaltonen:2011kc},
\begin{align}
& A_\text{FB}^\text{exp} =  0.158 \pm 0.075 && \text{(inclusive)} \,, \notag \\
& A_\text{FB}^\text{exp} =  0.475 \pm 0.114 && (m_{t\bar t} > 450~\text{GeV}) \,.
\label{ec:afbX}
\end{align}
the one for high $t \bar t$ invariant masses which exhibits the largest deviation ($3.4\sigma$) with respect to the SM prediction.
The total cross section at LHC is evaluated including four-fermion operators in a similar way. In order to display the effect of new contributions on the $t \bar t$ tail we evaluate the cross section for $t \bar t$ invariant masses larger than 1 TeV. 
We note that our calculation in terms of effective operators gives a larger (smaller) tail than the exact calculation for $t$-channel ($s$-channel) resonances. In the former case the differences are not dramatic but in the latter our results can be quite conservative, depending on the mass and width of the new resonance. A detailed comparison is presented in appendix~\ref{sec:b}.

\begin{figure}[htb]
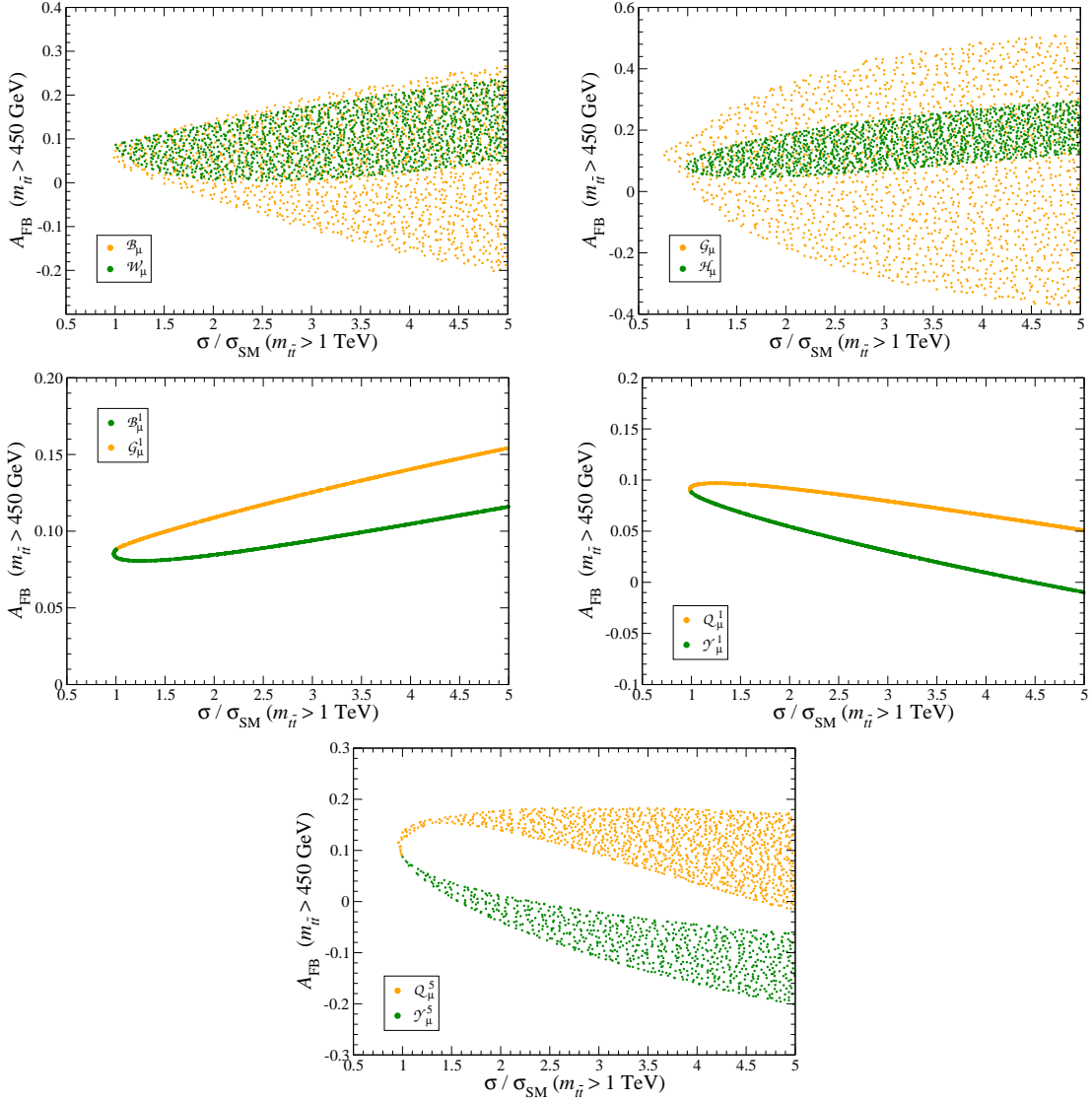

\begin{center}
\begin{tabular}{ccc}
\epsfig{file=Figs/bmu,height=4.8cm,clip=} & \quad &
\epsfig{file=Figs/gmu,height=4.8cm,clip=} \\
\epsfig{file=Figs/bg1,height=4.8cm,clip=} & \quad &
\epsfig{file=Figs/QY1,height=4.8cm,clip=} \\
\multicolumn{3}{c}{\epsfig{file=Figs/QY5,height=4.8cm,clip=}}
\end{tabular}
\caption{Allowed regions for the Tevatron $t \bar t$ asymmetry and the $t \bar t$ tail at LHC for a single vector boson in each representation.}
\label{fig:VB}
\end{center}
\end{figure}

The relation between the predictions for the Tevatron $t \bar t$ asymmetry and the $t \bar t$ tail at LHC is tested by performing a random scan over the relevant couplings $g_{ij}$ corresponding to each new particle or multiplet. The results for the ten vector boson representations are presented in Fig.~\ref{fig:VB}.
(For $\bmu^1$, $\gmu^1$, $\qmuo$ and $\ymuo$ the regions are one-dimensional because there is only one coupling involved.) There are several interesting conclusions which can be drawn from these plots:
\begin{enumerate}
\item For $\wmu$ and $\hmu$ the allowed regions are inside the corresponding ones for $\bmu$, $\gmu$, respectively. This is expected because the interactions of the former correspond to a particular case of the latter, with only left-handed couplings.
\item For new colour-singlet neutral bosons $\bmu$, $\wmu$ the linear terms have negative coefficients and decrease $\afb$, which can only reach the experimental value for large couplings when quadratic terms dominate. Hence,
accommodating a large asymmetry automatically implies a large $t \bar t$ tail.
For example, for $\afb \gtrsim 0.3$ the enhancement is more than a factor of five. This implies that these possible explanations for $\afb$ can be probed, and eventually excluded, with the luminosity collected in the 2010 run of LHC. The same conclusion applies to the vector bosons $\gmu^1$ and $\bmu^1$, as they only contribute in $d \bar d$ initial states and require a huge coupling to produce a large asymmetry.
\item For colour-octet isosinglet bosons $\gmu$ it is possible to have a large $\afb$ and still a moderate tail at LHC. This model provides an example of cancellation of linear $1/\Lambda^2$ terms in the cross section, provided that $g_{ii}^q = - g_{ii}^u = - g_{ii}^d$, {\em i.e.} the vector boson couples as an axigluon. However, in order to have positive coefficients in the interference terms the couplings for the third and first generation must have opposite sign. This does not happen for the isotriplet boson $\hmu$ which only has left-handed couplings, and for these SM extensions the predicted $t \bar t$ tail is large.
\item Another interesting candidate is a colour triplet $\qmu$, which has positive coefficients in interference terms as well. Its inclusion gives some enhancement to the asymmetry, which can reach the experimental value with the further addition of a scalar. Note that quadratic terms from operators $O_{qu^{(')}}$ decrease the asymmetry, as it can be derived from Eqs.~(\ref{ec:quad}) and is clearly seen in Fig.~\ref{fig:VB}. Hence, this model is interesting only for moderate couplings.
\item The rest of vector bosons, $\qmuo$, $\ymuo$ and $\ymu$ have little interest for the $t \bar t$ asymmetry because they do not allow for a value appreciably larger than the SM prediction.
\end{enumerate}
\begin{figure}[t]
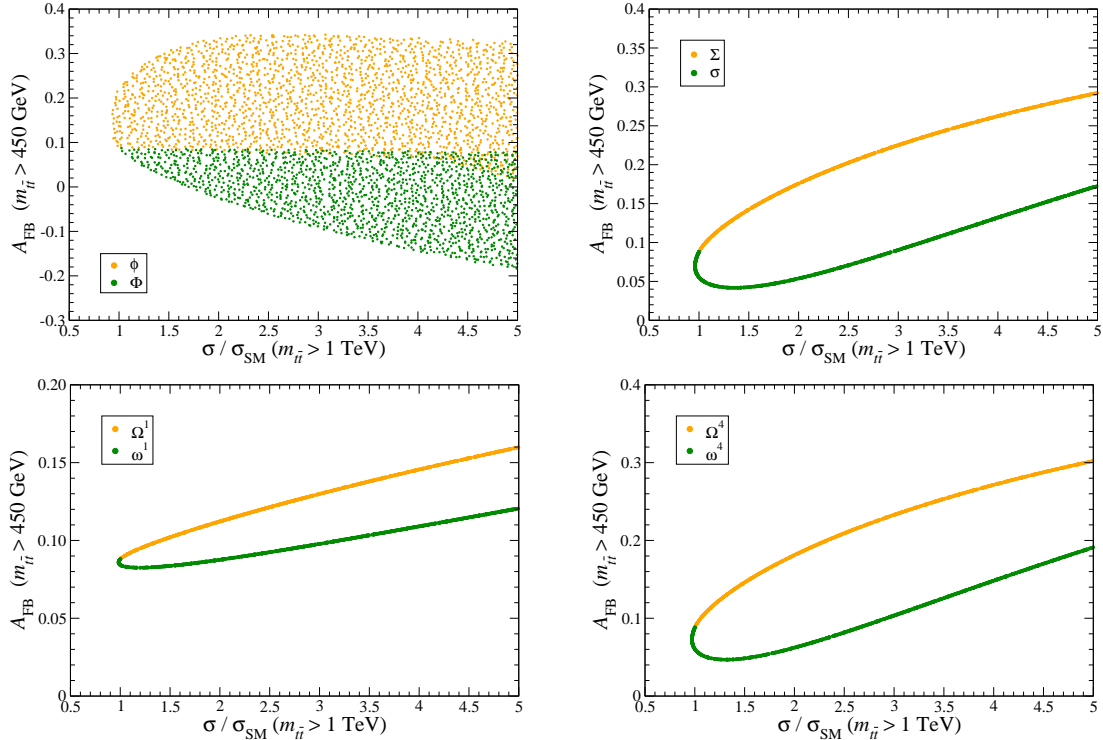

\begin{center}
\begin{tabular}{ccc}
\epsfig{file=Figs/phi,height=4.8cm,clip=} & \quad &
\epsfig{file=Figs/sigma,height=4.8cm,clip=} \\
\epsfig{file=Figs/Om1,height=4.8cm,clip=} & \quad &
\epsfig{file=Figs/Om4,height=4.8cm,clip=} 
\end{tabular}
\caption{Allowed regions for the Tevatron $t \bar t$ asymmetry and the $t \bar t$ tail at LHC for a single scalar in each representation.}
\label{fig:sc}
\end{center}
\end{figure}
Aside from these remarks, we also note that for (i) $\bmu^1$ and $\gmu^1$; (ii) $\qmuo$ and $\ymuo$; (iii) $\qmu$ and $\ymu$, the linear $1/\Lambda^2$ terms have opposite sign, which explains the behaviour observed in the plots for these vector bosons.

The results for the eight scalar representations are presented in Fig.~\ref{fig:sc}.
Except for the isodoublets $\phi$, $\Phi$, the regions are one-dimensional because there is only one coupling involved. We point out that:
\begin{enumerate}
\item A colour-singlet isodoublet $\phi$ (with the same quantum numbers as the Higgs boson) can give an asymmetry compatible with the experimental value, and still produce a moderate tail at LHC. (Note that quadratic terms involving the operators $O_{qu^{(')}}$, $O_{qd^{(')}}$ decrease the asymmetry.) On the other hand, a colour octet $\Phi$ produces an asymmetry smaller than the SM value because the interference terms have opposite sign.
\item For colour-sextets $\Omega^4$ and $\Sigma$ the interference terms increase the asymmetry because the operator coefficients are positive.\footnote{In Ref.~\cite{Shu:2009xf} the SM and colour-sextet contributions have a wrong relative sign, resulting in a decrease of the cross section at first order. The sign has been corrected in Refs.~\cite{Arhrib:2009hu,Ligeti:2011vt}.}
However, producing an asymmetry $\afb \gtrsim 0.3$ requires large couplings $g_{13}$ and implies a large $t \bar t$ tail at LHC, which might be already excluded. For the colour triplets $\omega^4$, $\sigma$ the situation is worse because the interference terms have negative operator coefficients, and even larger couplings are required to produce a large asymmetry.
\item For the two scalars $\Omega^1$, $\omega^1$ which only contribute in $d \bar d \to t \bar t$, the deviations in $\afb$ are always very small.
\end{enumerate}
To conclude this survey, we remark again that this correlation between $\afb$ and the $t \bar t$ tail at LHC applies to SM extensions with a single vector boson or scalar (as many of the ones proposed in the literature) but when more than one particle is present the contributions can add up or cancel, making it easier to fit the experimental data
and predict moderate effects in the $t \bar t$ tail, as we discuss in detail in the next section.

\section{A large $t \bar t$ asymmetry with a small $t \bar t$ tail}
\label{sec:3}

By inspection of Eqs.~(\ref{ec:int}) it is clear that the cancellation of the linear $1/\Lambda^2$ contributions to the cross section takes place provided
\begin{align}
& C_{qq'}^{1133} + C_{qq}^{3113} + C_{uu}^{3113} = C_{qu}^{1331} + C_{qu}^{3113} \equiv c_1 \,, \notag \\
& C_{qq'}^{1133} + 2 C_{ud'}^{3311} = C_{qu}^{1331} + C_{qd}^{3113} \equiv c_2
\label{ec:cancel}
\end{align}
(see also Ref.~\cite{Degrande:2010kt}).
Notice that in the left-hand side of both equations we have $LL$ and $RR$ couplings, whereas on the right-hand side we have $LR$ and $RL$ ones. As we have mentioned, one simple example where both equalities are fulfilled is an axigluon with flavour-diagonal couplings $g_{ii}^q = -g_{ii}^u = -g_{ii}^d$, with the additional requirement that first and third generation couplings have opposite sign, to have positive coefficients. (This model may be excluded by low-energy measurements, however~\cite{Chivukula:2010fk}.) But there are many other possibilities which can be constructed combining particles in Table~\ref{tab:CAint}, for instance, a colour triplet $\qmu$ together with a colour sextet $\Omega^4$ or $\Sigma$.

In these SM extensions with vanishing (or very small) contributions to the total cross section at first order, the FB asymmetry is
\begin{equation}
\afb = \afb^\text{SM} + \frac{2 c_1 (D_\text{int}^F-D_\text{int}^B)_{u \bar u}
+ 2 c_2 (D_\text{int}^F-D_\text{int}^B)_{d \bar d}}{\sigma_\text{SM}}
\end{equation}
plus smaller corrections from quadratic terms, which depend on the specific operators which yield $c_1$ and $c_2$.
Remarkably, one can obtain a good fit to both asymmetry measurements in Eqs.~(\ref{ec:afbX}) with values of $c_1$, $c_2$ of order unity.\footnote{With two parameters at hand we can fit the exact central values of the two measurements, but this requires huge values of the constants $c_1$, $c_2$.} For instance, assuming $c_1=c_2$ and equal $LL$, $RR$, $LR$ and $RL$ terms, the best fit to both measurements is $c_{1,2}=2$, for which
\begin{align}
& A_\text{FB}^{4F} =  0.225 && \text{(inclusive)} \,, \notag \\
& A_\text{FB}^{4F} =  0.366 && (m_{t\bar t} > 450~\text{GeV})
\label{ec:afb4F}
\end{align}
including linear and quadratic terms, with a $\chi^2$ of 1.72. For $c_2=0$ the best fit is found for $c_1 = 2.34$, giving similar predictions for the asymmetry.

\begin{figure}[t]
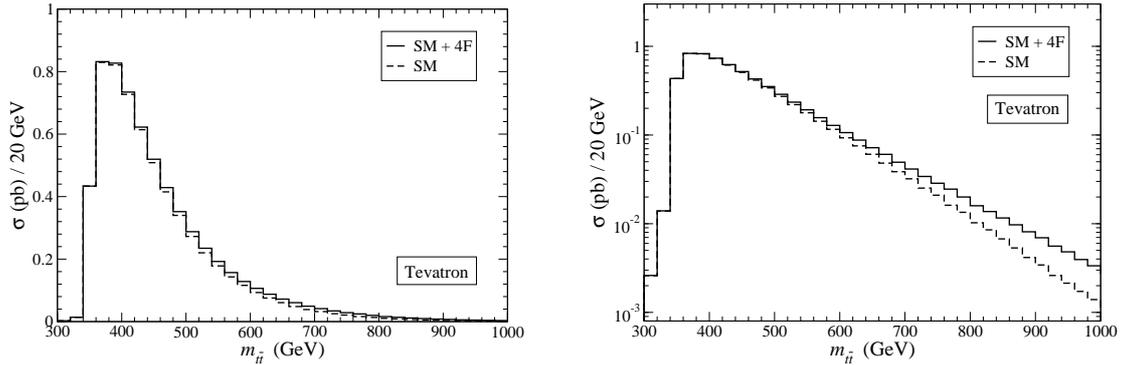

\begin{center}
\begin{tabular}{ccc}
\epsfig{file=Figs/mtt-tev,height=4.8cm,clip=} & \quad &
\epsfig{file=Figs/mttL-tev,height=4.8cm,clip=}
\end{tabular}
\caption{Invariant mass distribution for $t \bar t$ pairs at Tevatron, for the SM and with four-fermion contributions predicting the FB asymmetries in Eqs.~(\ref{ec:afb4F}). The plot on the left panel has linear scale whereas for the one in the right panel it is logarithmic.}
\label{fig:mtt-tev}
\end{center}
\end{figure}

We have investigated the effects on the $t \bar t$ tails by implementing four-fermion operators in the generator {\tt Protos}~\cite{AguilarSaavedra:2008gt}. We have first checked that this class of SM extensions does not produce a too large tail at Tevatron. Figure~\ref{fig:mtt-tev} shows the $t \bar t$ invariant mass distributions for the SM and with $LL+RR$, $RL+LR$ four-fermion contributions corresponding to $c_1=c_2=2$, which yield the asymmetries in Eqs.~(\ref{ec:afb4F}). Above $m_{t \bar t} = 700$ GeV the cross section is enhanced by $+56\%$, which is consistent with data~\cite{Aaltonen:2011kc}.
\begin{figure}[t]
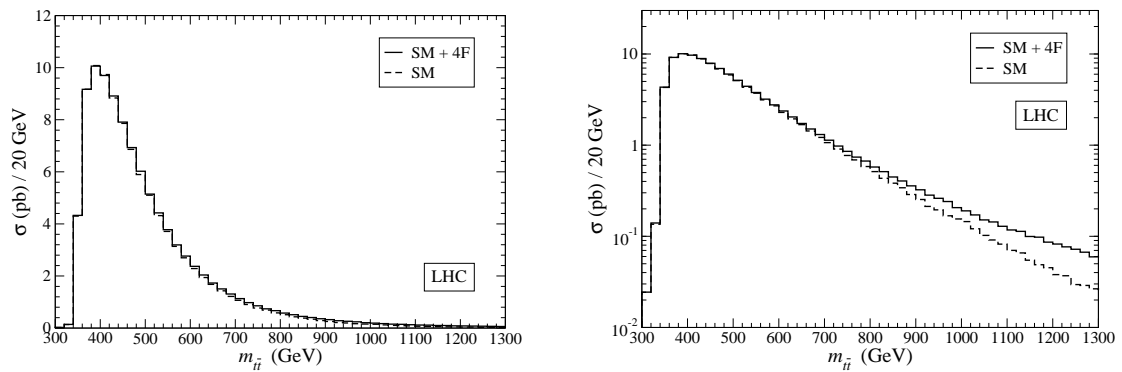

\begin{center}
\begin{tabular}{ccc}
\epsfig{file=Figs/mtt-lhc,height=4.8cm,clip=} & \quad &
\epsfig{file=Figs/mttL-lhc,height=4.8cm,clip=}
\end{tabular}
\caption{Invariant mass distribution for $t \bar t$ pairs at LHC, for the SM and with four-fermion contributions predicting the FB asymmetries in Eqs.~(\ref{ec:afb4F}). The plot on the left panel has linear scale whereas for the one in the right panel it is logarithmic.}
\label{fig:mtt-lhc}
\end{center}
\end{figure}
For LHC the invariant mass distributions are presented in Fig.~\ref{fig:mtt-lhc}. The cross section above 1 TeV is a factor of $2.3$ above the SM one. This deviation could be visible with the luminosity to be collected in 2011, provided that the systematic uncertainties (jet energy scale, jet energy resolution, $b$ jet energy scale, etc.) are low enough.
Here it is necessary to mention that for Tevatron a smaller efficiency for $t$-channel new physics at high $m_{t \bar t}$ has been recently claimed~\cite{Gresham:2011pa,Jung:2011zv}, which could help maintain the cross section at the high $m_{t \bar t}$ bins in agreement with measurements while reproducing the FB asymmetry. For LHC the efficiency decrease at $m_{t \bar t}>1$ TeV is not significant because of the larger detector coverage up to $|\eta| = 2.5$ for charged leptons and $|\eta| = 5$ for jets (see appendix~\ref{sec:b}).

\section{Conclusions}
\label{sec:4}

If the $t \bar t$ asymmetry measured at Tevatron corresponds to new physics, this new physics should also manifest at the $t \bar t$ tail at LHC. The size of the effect of course depends on the new physics itself which gives rise to the FB asymmetry, and it can serve to discriminate among different explanations. These issues have been investigated here using an effective operator framework and classifying all possible new vector bosons and scalars by their transformation properties under the SM gauge group. Particular models in the literature attempting to explain the observed asymmetry often fall into one of these classes.

For models which reproduce $\afb$ with $t$- and/or $s$-channel $Z'$ exchange~\cite{Jung:2009jz,Cao:2009uz,Choudhury:2010cd,Cao:2011ew,
Berger:2011ua,Gresham:2011pa}
we have found that the tail above 1 TeV should be enhanced by a factor of five at LHC,  at least for $Z'$ bosons heavier than 1 TeV. With $\sigma(m_{t \bar t} > 1~\text{TeV}) = 1.22$ pb at the tree level and a luminosity of approximately 35 pb$^{-1}$ collected in the 2010 run, the SM predicts 6.3 events in the semileptonic decay channel, not including the detector acceptance nor efficiencies. Then, it seems likely that an enhancement by a factor of five in the tail could be excluded just by analysing 2010 data. The same argument applies to $t$-channel $W'$ exchange~\cite{Cheung:2009ch,Cao:2010zb,Cheung:2011qa,Gresham:2011pa} or a mixture of both~\cite{Barger:2011ih}.

More exotic models explain the observed asymmetry by the exchange of a colour-triplet isosinglet scalar $\omega^4$ (see for example Refs.~\cite{Shu:2009xf,Arhrib:2009hu,Choudhury:2010cd,Ligeti:2011vt,Gresham:2011pa}) or its colour-sextet counterpart $\Omega^4$~\cite{Shu:2009xf,Arhrib:2009hu,
Grinstein:2011yv,Patel:2011eh,Ligeti:2011vt,Gresham:2011pa}.
These models could also be tested, and eventually excluded, with data already analysed by CMS and ATLAS in the search for $t \bar t$ resonances.
On the other hand, models with axigluons~\cite{Ferrario:2008wm,Frampton:2009rk,Cao:2010zb,Bai:2011ed,Choudhury:2010cd} or other types of colour-octet bosons~\cite{Djouadi:2009nb,Chen:2010hm,Burdman:2010gr,
Alvarez:2010js,Grinstein:2011yv,Gresham:2011pa} can in principle accommodate the measured $t \bar t$ asymmetry predicting a moderate $t \bar t$ tail at LHC. (Our discussion obviously does not directly apply to models where the
new physics produces $t \bar t$ plus other particles in the final state, see for example Ref.~\cite{Isidori:2011dp}.)
Besides, we note that Ref.~\cite{Delaunay:2011gv} has recently found that SM extensions explaining the FB asymmetry without predicting too large $t \bar t$ tails must have interference with the SM amplitudes.\footnote{This reference has appeared in the arXiv one day before the present paper, and our findings are consistent with theirs, where they overlap.} As we have shown, our conclusions are stronger because in many extensions with interference this is not possible either. Moreover, for all vector bosons and scalars in Table~\ref{tab:lagr} there is interference unless the involved couplings vanish.

Finally, in this paper we have investigated the conditions under which the first order $1/\Lambda^2$ contributions to the total $t \bar t$ cross section cancel while still producing a FB asymmetry compatible with experimental data. (A previous study at this order has been presented in Ref.~\cite{Degrande:2010kt}.)
Clearly, in this situation the tails of the $t \bar t$ invariant mass distribution are much smaller, both at Tevatron and LHC, as it has been shown explictly. A popular example of a model fulfilling these conditions is an axigluon with opposite couplings to the first and third generation, but there are many other possibilities which can be worked out from Table~\ref{tab:CAint}. All these SM extensions can accommodate the Tevatron $t \bar t$ asymmetry and cross section with small couplings, and predict a moderate enhancement of the $t \bar t$ tail at Tevatron and LHC, which is not in contradiction with experiment.
Interestingly, a small excess in the $t \bar t$ tail with boosted tops has been already observed at Tevatron~\cite{cdf10234}. In any case, these possible departures will soon be tested with forthcoming LHC data.

\section*{Acknowledgements}

This work has been partially supported by projects FPA2010-17915 (MICINN), FQM 101 and FQM 437 (Junta de Andaluc\'{\i}a) and CERN/FP/116397/2010 (FCT).

\appendix
\section{Four-fermion operators in $t \bar t$ production}
\label{sec:a}

We use the minimal basis in Ref.~\cite{AguilarSaavedra:2010zi} for gauge-invariant four-fermion operators. Fermion fields are ordered according to their spinorial index contraction, and
subindices $a$, $b$ indicate the pairs with colour indices contracted, if this pairing is different from the one for the spinorial contraction. Our basis consists of the following operators:

\noindent
(i) $\bar L L \bar L L$ operators
\begin{align}
& O_{qq}^{ijkl} = \oh (\bar q_{Li} \gM q_{Lj}) (\bar q_{Lk} \gm q_{Ll}) \,,
&& O_{qq'}^{ijkl} = \oh (\bar q_{Lia} \gM q_{Ljb}) (\bar q_{Lkb} \gm q_{Lla}) \,,
\notag \\
& O_{\ell q}^{ijkl} = (\bar \ell_{Li} \gM \ell_{Lj}) (\bar q_{Lk} \gm q_{Ll}) \,,
&& O_{\ell q'}^{ijkl} = (\bar \ell_{Li} \gM q_{Lj}) (\bar q_{Lk} \gm \ell_{Ll}) \,,
\notag \\
& O_{\ell \ell}^{ijkl} = \oh (\bar \ell_{Li} \gM \ell_{Lj}) (\bar \ell_{Lk} \gm \ell_{Ll}) \,.
\label{ec:LLLL}
\end{align}
(ii) $\bar R R \bar R R$ operators
\begin{align}
& O_{uu}^{ijkl} = \oh (\bar u_{Ri} \gM u_{Rj}) (\bar u_{Rk} \gm u_{Rl}) \,,
&& O_{dd}^{ijkl} = \oh (\bar d_{Ri} \gM d_{Rj}) (\bar d_{Rk} \gm d_{Rl}) \,,
\notag \\
& O_{ud}^{ijkl} = (\bar u_{Ri} \gM u_{Rj}) (\bar d_{Rk} \gm d_{Rl}) \,,
&& O_{ud'}^{ijkl} = (\bar u_{Ria} \gM u_{Rjb}) (\bar d_{Rkb} \gm d_{Rla}) \,,
\notag \\
& O_{eu}^{ijkl} = (\bar e_{Ri} \gM e_{Rj}) (\bar u_{Rk} \gm u_{Rl}) \,, 
&& O_{ed}^{ijkl} = (\bar e_{Ri} \gM e_{Rj}) (\bar d_{Rk} \gm d_{Rl}) \,,
\notag \\
& O_{ee}^{ijkl} = \oh (\bar e_{Ri} \gM e_{Rj}) (\bar e_{Rk} \gm e_{Rl}) \,.
\label{ec:RRRR}
\end{align}
(iii) $\bar L R \bar R L$ operators
\begin{align}
& O_{qu}^{ijkl} = (\bar q_{Li} u_{Rj}) (\bar u_{Rk} q_{Ll}) \,,
&& O_{qu'}^{ijkl} = (\bar q_{Lia} u_{Rjb}) (\bar u_{Rkb} q_{Lla}) \,,
\notag \\
& O_{qd}^{ijkl} = (\bar q_{Li} d_{Rj}) (\bar d_{Rk} q_{Ll}) \,,
&& O_{qd'}^{ijkl} = (\bar q_{Lia} d_{Rjb}) (\bar d_{Rkb} q_{Lla}) \,,
\notag \\
& O_{\ell u}^{ijkl} = (\bar \ell_{Li} u_{Rj}) (\bar u_{Rk} \ell_{Ll}) \,,
&& O_{\ell d}^{ijkl} = (\bar \ell_{Li} d_{Rj}) (\bar d_{Rk} \ell_{Ll}) \,,
\notag \\
& O_{qe}^{ijkl} = (\bar q_{Li} e_{Rj}) (\bar e_{Rk} q_{Ll}) \,,
&& O_{qde}^{ijkl} = (\bar \ell_{Li} e_{Rj}) (\bar d_{Rk} q_{Ll}) \,,
\notag \\
& O_{\ell e}^{ijkl} = (\bar \ell_{Li} e_{Rj}) (\bar e_{Rk} \ell_{Ll}) \,.
\label{ec:LRRL}
\end{align}
(iv) $\bar L R \bar L R$ operators
\begin{align}
& O_{qq\epsilon}^{ijkl} = (\bar q_{Li} u_{Rj}) \left[ (\bar q_{Lk} \epsilon)^T d_{Rl} \right] \,,
&& O_{qq\epsilon'}^{ijkl} = (\bar q_{Lia} u_{Rjb}) \left[ (\bar q_{Lkb} \epsilon)^T d_{Rla} \right] \,,
\notag \\
& O_{\ell q\epsilon}^{ijkl} = (\bar \ell_{Li} e_{Rj}) \left[ (\bar q_{Lk} \epsilon)^T u_{Rl} \right] \,,
&& O_{q \ell \epsilon}^{ijkl} = (\bar q_{Li} e_{Rj}) \left[ (\bar \ell_{Lk} \epsilon)^T u_{Rl} \right] \,.
\label{ec:LRLR}
\end{align}

In the calculation of the FB asymmetry, the interference of four-fermion and the tree-level SM contributions are
\begin{eqnarray}
\delta \sigma^{F,B}_\text{int}(u \bar u) & = & \frac{D_\text{int}^{F,B}}{\Lambda^2}
\left[ \Cqqp^{1133} + \Cqq^{3113} + \Cuu^{3113} \right]
- \frac{\tilde D_\text{int}^{F,B}}{\Lambda^2}
\left[ \Cqu^{1331} + \Cqu^{3113}  \right] \notag \,, \notag \\
\delta \sigma^{F,B}_\text{int}(d \bar d) & = & \frac{D_\text{int}^{F,B}}{\Lambda^2} \left[ \Cqqp^{1133}+2\,\Cudp^{3311} \right]
- \frac{\tilde D_\text{int}^{F,B}}{\Lambda^2} \left[ \Cqu^{1331}+\Cqd^{3113}  \right] \,,
\label{ec:int}
\end{eqnarray}
with the $D_\text{int}$ numerical coefficients satisfy $D_\text{int}^F = \tilde D_\text{int}^B$, $D_\text{int}^B = \tilde D_\text{int}^F$. They are collected in Table~\ref{tab:D}, evaluated for $m_t = 172.5$ GeV using CTEQ6L1 parton distribution functions~\cite{Pumplin:2002vw} with $Q=m_t$. An important remark to guide model building is that, since $D_\text{int}^F > D_\text{int}^B$, positive operator coefficients $C_x$ increase $\afb$ at first order. The pure four-fermion contributions are
\begin{eqnarray}
\delta \sigma^{F,B}_\text{4F}(u \bar u) & = &  \frac{D_1^{F,B}}{\Lambda^4}
\left[ \pr(\Cqq^{1133}+\Cqqp^{3113},\Cqqp^{1133}+\Cqq^{3113})
+ \pr(\Cuu^{1133},\Cuu^{3113}) \right] \notag \\[1mm]
& & + \frac{\tilde D_1^{F,B}}{\Lambda^4}
\left[ \pr(\Cqup^{1331},\Cqu^{1331}) + \pr(\Cqup^{3113},\Cqu^{3113}) \right] 
+ \frac{D_2}{\Lambda^4} \,
\pr(\Cqup^{3311},\Cqu^{3311}) \notag \\
& & - \frac{D_4}{\Lambda^4}
\left[ \pr(\Cqq^{1133}+\Cqqp^{3113},\Cqup^{1331},\Cqqp^{1133}+\Cqq^{3113},\Cqu^{1331}) \right. \notag \\[1mm]
&& \left. + \pr(\Cqup^{3113},\Cuu^{1133},\Cqu^{3113},\Cuu^{3113}) \right] \,, \notag \\
\delta \sigma^{F,B}_\text{4F}(d \bar d) & = &  + \frac{D_1^{F,B}}{\Lambda^4}
\left[ \pr(\Cqq^{1133},\Cqqp^{1133}) + 4 \pr(\Cud^{3311},\Cudp^{3311}) \right] \notag \\[1mm]
& & + \frac{\tilde D_1^{F,B}}{\Lambda^4}
\left[ \pr(\Cqup^{1331},\Cqu^{1331}) + \pr(\Cqdp^{3113},\Cqd^{3113}) + \oh \pr(\Cqqep^{1331},\Cqqe^{1331}) \right] \notag \\
& & + \frac{D_2}{\Lambda^4}
\, \pr(\Cqqe^{3311},\Cqqep^{3311})
+ \frac{D_3^{F,B}}{\Lambda^4} \RE \pr(\Cqqe^{3311},\Cqqep^{1331},\Cqqep^{3311},\Cqqe^{1331})
 \notag \\[1mm]
& & - \frac{D_4}{\Lambda^4}
\left[\pr(\Cqq^{1133},\Cqup^{1331},\Cqqp^{1133},\Cqu^{1331}) + 2 \pr(\Cqdp^{3113},\Cud^{3311},\Cqd^{3113},\Cudp^{3311})
\right] \,, \notag \\
\label{ec:quad}
\end{eqnarray}
with $D_1^F = \tilde D_1^B$, $D_1^B = \tilde D_1^F$. We have used the functions
\begin{eqnarray}
\pr(x,y) & = & |x|^2 + |y|^2 + \frac{2}{3} \, \RE x y^* \,, \notag \\
\pr(x,y,u,v) & = & xy^* + uv^* + \frac{1}{3} xv^* + \frac{1}{3} uy^*
\end{eqnarray}
to write the expressions in a more compact form. The numerical coefficients are collected in Table~\ref{tab:D}. Interference of four-fermion corrections and SM NLO corrections are not considered. 
 
\begin{table}[htb]
\begin{center}
\begin{tabular}{lcccccccc}
 & $D_\text{int}^F$ & $D_\text{int}^B$
 & $D_1^F$ & $D_1^B$ & $D_2$
 & $D_3^F$ & $D_3^B$ & $D_4$ \\
$u \bar u$ inclusive & 0.522 & 0.228
  & 91.7 & 21.4 & 74.8 & -- & -- & 40.2 \\
$d \bar d$ inclusive & 0.0855 & 0.0409 
  & 12.6 & 3.32 & 10.25 & 5.63 & 14.9 & 6.64 \\
$u \bar u$ $m_{t\bar t} > 450~\text{GeV}$ & 0.318 & 0.108
  & 74.5 & 15.0 & 61.1 & -- & -- & 24.24 \\
$d \bar d$ $m_{t\bar t} > 450~\text{GeV}$ & 0.0443 & 0.0161
  & 9.15 & 1.98 & 7.50 & 3.91 & 11.1 & 3.39
\end{tabular}
\end{center}
\caption{Numerical coefficients for interference and four-fermion contributions to the $t \bar t$ asymmetry. The units of $D_\text{int}^{F,B}$ are $\text{pb} \cdot \text{TeV}^2$ and the units of $D_i^{F,B}$ are $\text{fb} \cdot \text{TeV}^4$.}
\label{tab:D}
\end{table}

The $t \bar t$ cross section at LHC is evaluated using analogous expressions but different numerical constants, covering the forward and backward hemispheres. Because we are interested in the relative enhancement of the high $m_{t \bar t}$ tail, we use tree-level calculations everywhere to be consistent. The tree-level SM cross section (including all subprocesses) is 1.22 pb, and the four-fermion operator contributions to $u \bar u,d \bar d \to t \bar t$ are determined by the coefficients in Table~\ref{tab:D7}.

\begin{table}[htb]
\begin{center}
\begin{tabular}{lccccc}
 & $D_\text{int}$ & $D_1$ & $D_2$ & $D_3$ & $D_4$ \\
$u \bar u$ $m_{t\bar t} > 1~\text{TeV}$ 
 & 240.5 & 315.6 & 465.9 & -- & 31.41 \\
$d \bar d$ $m_{t\bar t} > 1~\text{TeV}$ 
 & 129.2 & 159.6 & 235.4 & 235.8 & 16.85
\end{tabular}
\end{center}
\caption{Numerical coefficients for interference and four-fermion contributions to the $t \bar t$ cross section at LHC. The units of $D_\text{int}$ are $\text{fb} \cdot \text{TeV}^2$ and the units of $D_i$ are $\text{fb} \cdot \text{TeV}^4$.}
\label{tab:D7}
\end{table}

\section{Comparison with exact calculations}
\label{sec:b}

We test the range of validity of our effective operator approximation by comparing with exact results for  a $t$-channel $Z'$ in the representation $\bmu$ and an $s$-channel $g'$ in the representation $\gmu$. We plot the results of the four-fermion operator and exact calculations as a function of the new particle mass $M \equiv \Lambda$ keeping $C/\Lambda^2$ constant, so that the four-fermion predictions remain flat while the exact ones deviate from this limit for lower $\Lambda$. 
Note that, for example, for $m_{t \bar t} = \sqrt{\hat s} = 1$ TeV, $\hat t$ ranges from $-0.0012~\text{TeV}^2$ to $-0.94~\text{TeV}^2$.

\begin{figure}[htb]
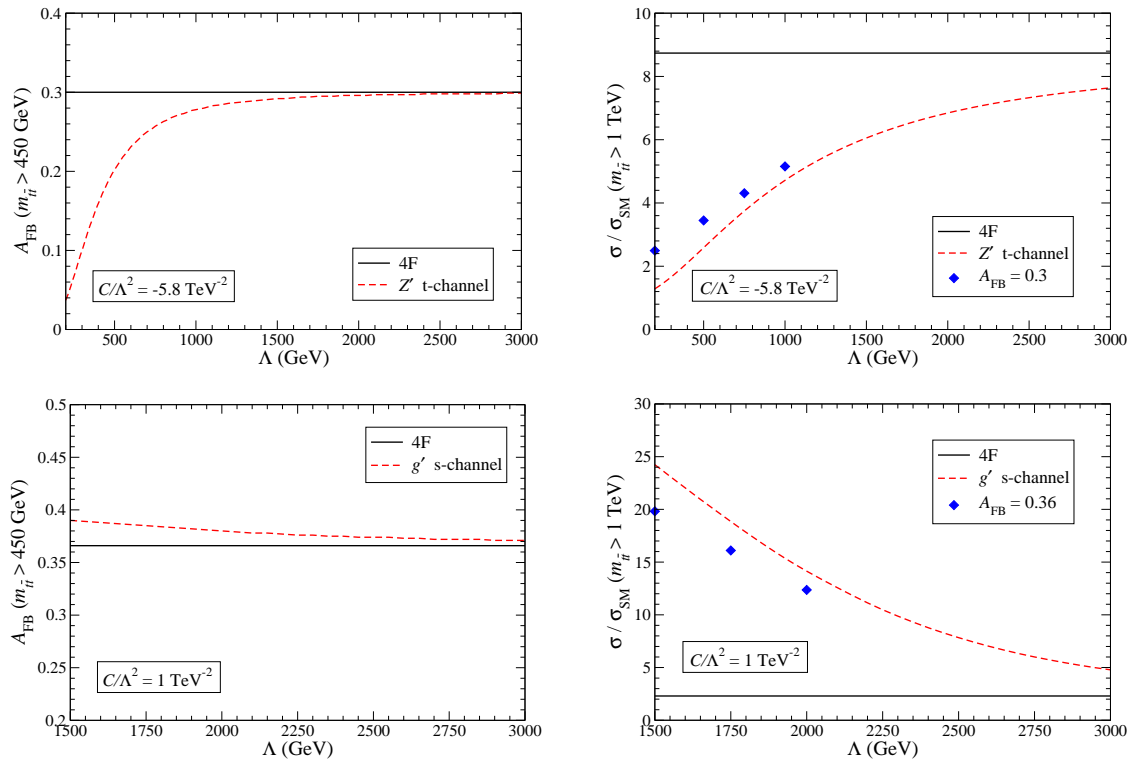

\begin{center}
\begin{tabular}{ccc}
\epsfig{file=Figs/afb-M,height=4.8cm,clip=} & \quad &
\epsfig{file=Figs/tail-M,height=4.8cm,clip=} \\[2mm]
\epsfig{file=Figs/afb2-M,height=4.8cm,clip=} & \quad &
\epsfig{file=Figs/tail2-M,height=4.8cm,clip=}
\end{tabular}
\caption{Comparison between four-fermion operator and exact calculations for $t$ and $s$ channels (see the text).}
\label{fig:comp}
\end{center}
\end{figure}

For a $t$-channel $Z'$ we select $C_{uu}^{3113}/\Lambda^2 = -5.8~\text{TeV}^{-2}$, which gives $\afb = 0.3$ in the effective operator approximation. We observe in Fig.~\ref{fig:comp} (up) that the exact calculation yields both a smaller asymmetry (left) and smaller tail $\sigma/\sigma_\text{SM}$ for $m_{t \bar t}>1~\text{TeV}$ (right). 
For small $\Lambda$ the coupling assumed is not large enough to generate the required asymmetry;  
for this reason we have also calculated the values of $\sigma/\sigma_\text{SM}$ for $Z'$ masses $M=200,500,750,1000$ GeV and larger couplings $C$ so as to reproduce $\afb = 0.3$. 
These four points are displayed in blue in Fig.~\ref{fig:comp} (up, right). From this analysis we can conclude that for $\Lambda > 1~\text{TeV}$ the effective operator formalism is accurate enough for our purposes. For $200~\text{GeV} < \Lambda < 1~\text{TeV}$ our calculations can overestimate the tail by up to a factor of $2-3$, but the actual increase in the cross section is still significant.

For an $s$-channel $g'$ we select $C/\Lambda^2 =1~\text{TeV}^{-2}$ as in the example of section~\ref{sec:3}, giving $\afb = 0.366$. We assume for $g'$ a large width $\Gamma = 0.1 M$,
and only consider masses above 1.5 TeV. The $t \bar t$ asymmetry with the exact calculation is slightly above the one obtained with the effective operator approximation and, as anticipated, the tail cross section $\sigma/\sigma_\text{SM}$ at LHC is up to one order of magnitude larger because of the $M/\Gamma$ propagator enhancement. We also display the value of $\sigma/\sigma_\text{SM}$ for three $g'$ masses $M=1.5,1.75,2$ TeV and the $C$ values which reproduce $\afb = 0.366$ in the exact calculation.
With this example we can confirm that in this case our effective operator calculations are quite conservative and the effects of $s$-channel resonances can be much larger, depending on their mass and width.

\begin{figure}[t]
\begin{center}
\epsfig{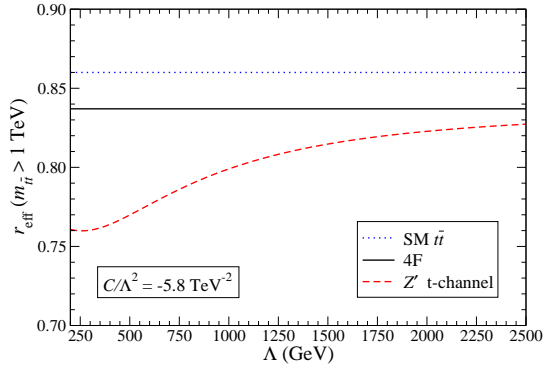}
\caption{Comparison between the charged lepton acceptance for new physics contributions from four-fermion operators and a $t$-channel $Z'$.}
\label{fig:comp2}
\end{center}
\end{figure}

Finally we perform a parton-level simulation including the $t \bar t$ pair decay to investigate a possible efficiency decrease at high $m_{t \bar t}$ due to forward scattering by new $t$-channel resonances~\cite{Gresham:2011pa,Jung:2009jz}. The fraction of events in which the charged lepton is too forward or too soft to be detected can be measured by the ratio
\begin{equation}
r_\text{eff} = \frac{\delta \sigma'}{\delta \sigma^\text{sl}} = \frac{\sigma'-\sigma'_\text{SM}}{\sigma^\text{sl}-\sigma^\text{sl}_\text{SM}} ~,\quad m_{t \bar t} > 1~\text{TeV}
\end{equation}
where $\sigma^\text{sl}$ are the total cross sections times semileptonic branching ratio and the primed quantities are the same but also requiring for the charged lepton a pseudo-rapidity $|\eta| < 2.5$ and transverse momentum $p_T > 20$ GeV. We do not include jet acceptances since the rapidity coverage is much larger than for leptons, $|\eta|<5$. Keeping $C_{uu}^{3113}/\Lambda^2 = -5.8~\text{TeV}^{-2}$ constant, we plot $r_\text{eff}$ in Fig.~\ref{fig:comp2} for the new physics contributions (four-fermion operators and $t$-channel $Z'$) as well as for the SM. It is clear that the acceptance decrease due to forward scattering is unimportant even for very small resonance masses due to the good rapidity coverage of the CMS and ATLAS detectors. (This is in agreement with calculations in Ref.~\cite{Jung:2011zv}.) Nevertheless, for high $t \bar t$ invariant masses the reconstruction of boosted top quarks may have to be optimised in order to achieve a large overall efficiency.

\end{document}